\documentstyle[aps]{revtex}

\def\<{\langle}
\def\>{\rangle}
\def\b{\begin{equation}}
\def\e{\end{equation}}
\def\be{\begin{eqnarray}}
\def\ee{\end{eqnarray}}

\begin{document}
\title{On the mean value of the energy for resonant states}
\author{{O. Civitarese $^{a)}$, M. Gadella $^{b)}$ and R. Id Betan $^{c)}$}}
\address{{\it {$^{a)}$ Dept. of Physics, Univ. of La Plata, C.C.67 (1900) La Plata.}}%
\\
Argentina\\
{\it {$^{b)}$ Dept. of Theoretical Physics, Univ. of Valladolid, Valladolid.}%
}\\
Spain\\
{\it {$^{c)}$ Dept. of Physics, Univ. of Rosario, Rosario. Argentina}}}
\maketitle

\begin{abstract}
In this work we discuss possible definitions of the mean value of the energy
for a resonant (Gamow) state. The mathematical and physical aspects of the
formalism are reviewed. The concept of rigged Hilbert space is used as a
supportive tool in dealing with Gamow-resonances.

\end{abstract}

\section{Introduction}

The use of resonant(Gamow) states in nuclear structure calculations was
proposed years ago by the Stockholm-Debrecen group \cite{liotta96}\cite{REP}
and since then the notion has been widely applied to a variety of physical
situations, with a remarkable success. The reader is kindly referred to
Refs. \cite{liotta96}\cite{REP} and references therein for a comprehensive
introduction on the subject. Although Gamow's idea of decaying states was
immediately recognized as a major breakthrough in Quantum Mechanics \cite
{GAMOW}, its use in modern nuclear structure calculations was delayed for
nearly forty years until the work of Tore Berggren \cite{TORE} shows that
single-particle basis (Beggren's basis) can accommodate single particle
resonant states of complex energy. Berggren's suggestions were adopted by
Liotta and co-workers \cite{LIOTTA} thus given structural identity to a
modern view of the continuum and its effects upon nuclear structure
observables \cite{REP}. Mathematically speaking the Stockholm approach is,
as a matter of fact, based on the identification of single-particle
resonances, in the standard one-body nuclear central potential \cite{VERTSE}%
, and in the numerical calculation of the needed matrix elements of nuclear
two-body interactions \cite{LIOTTA}. So far, the numerical treatment of
resonant states, either by performing explicitly the needed integrals in the
complex plane \cite{TORE}\cite{VERTSE} or by projecting them on the real
axis as non-overlapping states with a Breit-Wigner broadening, has proved
its feasibility. A parallel formal development on Gamow's resonances
followed from the work of Bohm and Gadella, which is documented in a
rigorous mathematical way in \cite{BOOKBG}. However, a link between these
two fronts of research on Gamow resonant states is missed. Particularly, we
have noticed that basic notions, as the one of the value of the expectation
value of the Hamiltonian on a resonant state, still need to be clarify
and/or discuss in detail. In this paper we focus on the question about the
definition of the expectation value of the Hamiltonian on a resonant state.
We shall illustrate this point as follows. In ordinary Quantum Mechanics the
mean value of the energy on a state represented by the density $\rho $ is
given by ${\rm {Tr}\,\rho H}$, where $H$ is the Hamiltonian. If $\rho $
represents a pure state it has the form $\rho =|\psi \rangle \langle \psi |$
and the quantity ${\rm {Tr}\,\rho H}$ gives the usual formula for the mean
value of the energy on the state $|\psi \rangle $, which is written as $%
\langle \psi |H|\psi \rangle $. Gamow resonant states (or Gamow states, for
short), are state vectors representing exponentially decaying states. They
are not ordinary quantum states. Such a quantum state has a Breit-Wigner
energy distribution, which is strictly nonzero for all values of the energy $%
E$, i.e. $-\infty <E<\infty $. This contradicts the fact that the spectrum
of the physical Hamiltonian $H$ should be lower-bound. Therefore, the Gamow
state cannot be represented as a vector in a Hilbert space in which $H$ is a
self-adjoint operator \cite{2}\cite{3}\cite{4}\cite{5}. In addition, if $%
\varphi $ is a Gamow state, it must fulfill a condition of the kind $%
e^{-itH}\varphi =Ae^{-\gamma t}\varphi $, $A$ being a phase ($A=e^{\phi (t)}$
with $\phi (t)$ real). Thus, $H\varphi =(\phi (t)/t+i\gamma )\varphi $ and
the Gamow vectors must be eigenvectors of the total Hamiltonian with complex
eigenvalues. Obviously this condition cannot be satisfy in the ordinary
Hilbert space, since $H$ must be self adjoint. In fact, Gamow states can be
properly defined as functionals on a certain space of test vectors, in the
same form that the generalized eigenstates of the position and momentum
operators are defined \cite{4}\cite{5}. The space of functionals contains
also the Hilbert space of ordinary states, so that no information is lost
from ordinary Quantum Mechanics. Thus, we have a triplet of spaces $\Phi
\subset {\cal H}\subset \Phi ^{\times }$, called the rigged Hilbert space or
Gelfand triplet.

Here ${\cal H}$ is the Hilbert space, $\Phi $ the space of test vectors and $%
\Phi ^{\times }$ its antidual\footnote{$F\in \Phi ^{\times }$ if it is a
mapping from $\Phi $ into the complex plane $C$ with the following
conditions (i) Antilinearity: $F(\alpha \varphi +\beta \phi )=\alpha
^{*}F(\varphi )+\beta ^{*}F(\phi )$, for all $\alpha ,\,\beta \,\in C$ and
all $\varphi ,\,\phi \in \Phi $. (ii) Continuity: $F$ must be a continuous
mapping. $\Phi ^{\times }$ is a topological vector space with a topology
which is coarser than the Hilbert space topology on ${\cal H}$.}. In this
context, Gamow states will exist in the dual space $\Phi ^{\times }$ \cite
{BOOKBG}\cite{7}. At this point a difficulty arises, namely: $\Phi ^{\times
} $ is not an inner product space. If we represent the Gamow vector as $%
|f_{0}\rangle $, the bracket $\langle f_{0}|H|f_{0}\rangle $ and the bracket 
$\langle f_{0}|f_{0}\rangle $ are not defined. Thus, in principle, we cannot
define the mean value of $H$ on $|f_{0}\rangle $ as we would do in ordinary
Quantum Mechanics. This is precisely the sort of questions which we meant
above.

The aim of the present paper is to recall on the attempts to define the mean
value of $H$ on a Gamow state and discuss their advantages or disadvantages.
These attempts use the following concepts:

1.- The mean value of the energy of a decaying state must be zero, because
the energy of a decaying process should be invariant. A non-zero energy will
be in contradiction with the principle of conservation of the energy \cite{8}%
\cite{9}\cite{10}.

2.- The energy average of a Gamow state should be complex because the Gamow
state is an eigenvalue of $H$ with complex energy.

3.- If we admit that Gamow states are genuine quantum states they must have
a real energy average that should be determined from first principles \cite
{TORE}\cite{4}\cite{12}.

4.- Gamow states admit a representation as {\bf normalizable} vectors in a
Hilbert space in which the Hamiltonian is {\bf not} a self adjoint operator
and it has an spectrum of eigenvalues extending from $-\infty$ to $\infty$.
In this case, we can define scalar products of Gamow vectors and a mean
value of the energy, which is real \cite{4}.

In the last section, we shall present and evaluate these four possibilities.
Before, in the next section, we recall the definition of a Gamow vector and
enumerate some of its properties.

\section{The Gamow vectors and their properties.}

In ordinary Quantum Mechanics, it is customary to associate a normalizable
vector (with unitary norm ) to each pure quantum state. If we have a quantum
decaying state and we want to assign to it a vector $\psi $ in a separable
Hilbert space we can define its {\it non-decay probability} for $t>0$ as 
\cite{4}\cite{5}\cite{13} 
\begin{equation}
P(t)=|A(t)|^{2},\,\,{\rm {where}\,\,}A(t)=<\psi |e^{-itH}|\psi >.  \label{1}
\end{equation}
Since $\psi $ is normalized one has $0\le P(t)\le 1$ and the decay
probability is given by $1-P(t)$. $A(t)$ is called the non-decay amplitude.
It has been shown that $A(t)$ is roughly exponential for values of time
which are not too short neither too long. At $t=0$ the time derivative of $%
A(t)$ is zero and this excludes the exponential time behaviour of $\psi $
for small times. For large values of time $A(t)$ goes to zero slower than an
exponential \cite{13}.

This is what the theory predicts {\it provided that a decaying state can be
represented by a vector in a Hilbert space}. We have already mentioned in
the Introduction another argument to show that an exponentially decaying
state cannot be represented by a vector in a Hilbert space.

The experimental evidence on decaying states shows a decay which seems to be
exponential up to the degree of experimental accuracy. In consequence, it is
natural to consider exponentially decaying vector states as true physical
states. But, we have to find out the mathematical nature of these objects.
This problem has been satisfactorily solved by Arno Bohm and coworkers, in
terms of rigged Hilbert spaces or Gelfand triplets \cite{BOOKBG}\cite{14}%
\cite{Bohm81}\cite{15}\cite{Gadella83}\cite{Gadella84}\cite{16}.

In fact, we can look at a decaying process as the second half of a resonant
scattering process in which the process of formation of a resonance is
ignored \cite{4}. Resonances can be defined in several ways, but it is
generally accepted that, under rather general conditions, we can associate a
resonance to each of the pairs of poles of the analytic continuation of the $%
S$-matrix, $S(E)$, on the real semi-axis of the energy values \cite{2}\cite
{3}\cite{4}\cite{13}\cite{17}. In many implementable physical cases we can
assume that in the resonant scattering process both the ``free'' Hamiltonian 
$H_{0}$ and the ``perturbed'' Hamiltonian $H=H_{0}+V$, where $V$ is a
suitable potential, have the same continuous spectrum. This spectrum
coincides with the positive semi-axis $R^{+}=[0,\infty )$ and the identity
of both continuous spectra is a consequence of the asymptotic completeness
of the scattering. This property implies that the Moller operators $\Omega
_{\pm }$ exist and fulfill the following relation \cite{BOOKBG} 
\begin{equation}
\Omega _{\pm }H_{0}=H\Omega _{\pm }\;\;{\rm or\;\;}H=\Omega _{\pm
}H_{0}\Omega _{\pm }^{\dagger }  \label{2}
\end{equation}
Furthermore, to simplify the formalism, we can assume that the energy
spectrum is non-degenerate (which is the case of a spherically symmetric
potential in the $l=0$ channel). In this case, it exists a unitary operator
which connects the abstract Hilbert space of states and $L^{2}(R^{+})$ (the
Hilbert space of the energy representation) such that $UH_{0}U^{-1}\phi
(E)=E\phi (E)$, where $U$ {\it diagonalizes} the free Hamiltonian $H_{0}$.

The explicit construction of the rigged Hilbert spaces for the decay process
is relevant in our discussion and, therefore, we want to summarize it here.
Further details can be found in the literature \cite{BOOKBG}\cite{7}\cite{15}%
\cite{Gadella83}\cite{Gadella84}. First, we take the spaces of the so called 
{\it very well behaved functions}. These spaces are defined by analytic
functions on the upper or the lower half planes of the complex plane $C$
that vanish at infinity. The boundary values of these functions on the
positive semi-axis $R^{+}$ are uniquely defined and, furthermore, it is
possible to recover all the values of these functions knowing their boundary
values on $R^{+}$ \cite{18}. Let us call $\Delta _{\pm }$ the spaces of
these boundary values. We can construct a rigged Hilbert space suitable for
the definition of decaying Gamow vector as follows. Firstly, we define $\Phi
_{+}=\Omega _{+}U^{-1}\Delta _{+}$, where $\Omega _{+}$ is the outgoing
Moller operator. Then, if ${\cal H}$ is the Hilbert space of scattering
states \footnote{%
I.e., the absolutely continuous space for the total Hamiltonian $H$.}, we
have the following rigged Hilbert space 
\begin{equation}
\Phi _{+}\subset {\cal H}\subset (\Phi _{+})^{\times }  \label{3}
\end{equation}

Since every function in $\Delta _{+}$ determines {\it uniquely} an analytic
function on the upper half plane for every vector $\varphi _{+}\in \Phi _{+}$
we have an analytic function $\phi (E)_{+}$ on the upper half plane. Its
complex conjugate $\phi _{+}^{\#}(E)=[\phi _{+}(E)]^{*}$ is an analytic
function on the {\it lower half plane}. Then, if $z_{R}=E_{R}-i\Gamma /2$
denotes a resonance pole ($\Gamma >0$) the Gamow vector $|f_{0}>$ can be
defined as a functional on $\Phi _{+}$ as\footnote{%
Here, we are using the notation in \cite{19}} 
\begin{equation}
<\varphi _{+}|f_{0}>=\phi _{+}^{\#}(z_{R})  \label{4}
\end{equation}

This definition implies several important properties namely: (i) $|f_{0}>\in
\Phi _{+}^{\times }$ is a continuous anti-linear functional on $\Phi _{+}$,
(ii) $|f_{0}>\notin {\cal H}$, (iii) the Hamiltonian $H$ satisfies $H\Phi
_{+}\subset \Phi _{+}$ and is continuous with the topology on $\Phi _{+}$.
It means that $H$ can be continuously extended (with the weak topology) to a
continuous operator on $\Phi _{+}^{\times }$, so that the action of $H$ on $%
|f_{0}>$ is a well defined operation. Then, we have $H|f_{0}>=z_{R}|f_{0}>$,
which means that the Gamow vector $|f_{0}>$ is an eigenvector of $H$ with
eigenvalue $z_{R}$. This is possible because $\Phi _{+}^{\times }$ is not a
Hilbert space, and, (iv) the time evolution of $|f_{0}>$ is only possible
for positive values of $t$. We can show that \cite{BOOKBG}\cite{7} 
\begin{equation}
e^{-itH}|f_{0}>=e^{-itE_{R}}\,e^{-\Gamma t}\,|f_{0}>  \label{5}
\end{equation}
This means that $|f_{0}>$ decays exponentially. All these properties justify
the choice of $|f_{0}>$ as a representation of the decaying Gamow vector.

In a resonant scattering process, together with the decaying channel, it
exists a process of {\it capture} or formation of a resonance. This is
called the growing or capture process and it can be described by the
evolution of another functional, which is the growing Gamow vector $|%
\widetilde{f}_{0}>$\footnote{%
The question about the physical meaning of this functional i.e.: whether it
is related to the capture or creation of a resonance, has not been answered
yet. If capture and decaying processes are not equally probable they can not
be symmetric in the sense presented here. Thus, in this interpretation, $|%
\tilde{f}_{0}>$ would be the time reversal of the Gamow vector $|f_{0}>$.}.
This is a functional on $\Phi _{-}$ and therefore an element of the vector
space $\Phi _{-}^{\times }$. As functional its definition coincides with the
one given in Eq. (\ref{4}) by replacing the function $\phi _{+}(E)$ by $\phi
_{-}(E)$, which is analytic on the lower half plane, and the point $z_{R}$
by its complex conjugate $z_{R}^{*}=E_{R}+i\Gamma /2$ 
\begin{equation}
<\varphi _{-}|\widetilde{f}_{0}>=\phi ^{\#}(z_{R}^{*}).  \eqnum{4'}
\label{4'}
\end{equation}
Its properties are the following: (i) $|\widetilde{f}_{0}>\notin {\cal H}$,
(ii) $H|\widetilde{f}_{0}>=z_{R}^{*}|\widetilde{f}_{0}>$ and, (iii) its time
evolution is well defined if $t<0$ 
\begin{equation}
e^{-itH}|\widetilde{f}_{0}>=e^{-itE_{R}}\,e^{\Gamma t}\,|\widetilde{f}_{0}>
\label{6}
\end{equation}
i.e., $|\widetilde{f}_{0}>$ increases exponentially until $t=0$, which is
conventionally the time at which the capture process is completed and the
decay starts.

From these definitions it can be concluded that Gamow vectors obey

\begin{equation}
<f_{0}|\varphi _{+}>=<\varphi _{+}|f_{0}>\;\;{\rm and}\;\;<\widetilde{f}%
_{0}|\varphi _{-}>=<\varphi _{-}|\widetilde{f}_{0}>^{*},  \label{7}
\end{equation}
where the star denotes complex conjugation. The bra vectors $<f_{0}|$ and $<%
\widetilde{f}_{0}|$ are continuous {\it linear} functionals on $\Phi _{+}$
and $\Phi _{-}$, respectively, 
\begin{equation}
<f_{0}|H=z_{R}^{*}<f_{0}|;\;\;\;<\widetilde{f}_{0}|H=z_{R}<\widetilde{f}%
_{0}|.  \label{8}
\end{equation}
The time evolution of $<f_{0}|$ is defined for $t>0$ only and it gives 
\begin{equation}
<f_{0}|e^{itH}=<f_{0}|e^{itE_{R}}\,e^{-\Gamma t}  \label{9}
\end{equation}
and the time evolution of $<\widetilde{f}_{0}|$ is defined for $t<0$ as 
\begin{equation}
<\widetilde{f}_{0}|e^{itH}=<\widetilde{f}_{0}|e^{itE_{R}}\,e^{\Gamma t}
\label{10}
\end{equation}

In this context, it seems that exponentially behaving state vectors are the
only class of vectors which can represent Gamow states. To construct a
representation, one has to consider these states and also a {\it background}%
, physically produced by the interaction with the environment, re-scattering
processes, etc \cite{13} and mathematically by contour integrals in the
complex plane \cite{4}\cite{14}\cite{BohmA81}\cite{15}\cite{Gadella83}\cite
{Gadella84}. Usually, the resonant scattering process is produced by the
interaction of a free prepared state with a potential, creating the
resonance. The prepared state must be represented by a Hilbert space vector,
say $\varphi $. Then, if $\Omega _{-}\varphi =\varphi _{-}$, we can show
that $\varphi _{-}$ is the sum of two contributions. One of these
contributions is proportional to the decaying Gamow vector $|f_{0}>$ and the
other to the background, represented by a vector in $\Phi _{+}^{\times }$,
so that \cite{4}\cite{14}\cite{BohmA81}\cite{15}\cite{Gadella83}\cite
{Gadella84} 
\begin{equation}
\varphi _{-}=a\,|f_{0}>+|{\rm background}>  \label{11}
\end{equation}
where $a$ is a complex number. This background would be responsible for the
deviations of the exponential law on the range of short and large times.

Our next goal is to present and compare the definitions of the energy
average on Gamow vectors.

\section{Definitions of energy average on Gamow vectors.}

In this section we shall review the known results obtained in dealing with
the definition of energy averages on Gamow states, which are available in
the literature, as well as our own definition of it.

\subsection{The mean value of the energy is equal to zero.}

It was Nakanishi who first proposed this idea \cite{8}. In fact, if $%
H|f_{0}>=z_{R}|f_{0}>$ and $<f_{0}|H=z_{R}^{*}<f_{0}|$, we have that $%
<f_{0}|H|f_{0}>=z_{R}<f_{0}|f_{0}>=z_{R}^{*}<f_{0}|f_{0}>$. This implies
that $(z_{R}-z_{R}^{*})<f_{0}|f_{0}>=0\Longrightarrow <f_{0}|f_{0}>=0$ and,
therefore, $<f_{0}|H|f_{0}>=0$. The weak point of this argument is that the
bracket $<f_{0}|f_{0}>$ is not defined. Further attempts to define it have
been made \cite{A}, but the results are not convincing from a mathematical
point of view.

\subsection{The averages are complex.}

In specific models, like Friedrichs's model, the bracket $<\widetilde{f}%
_{0}|f_{0}>$ is well defined and its value is one \cite{20}. If we try to
obtain this result in a general model independent setting we conclude that $<%
\widetilde{f}_{0}|f_{0}>$ can be defined as a distribution-kernel and that
it has the value one, although it is not clear if this is the unique choice 
\cite{21}. If we now define $\Pi =|f_{0}><\widetilde{f}_{0}|$, it is now
obvious that $\Pi ^{2}=\Pi $. This idempotency suggest us that that $\Pi $
could be taken as the density operator for the decaying Gamow vector $%
|f_{0}> $. Now if it would be possible to define {\rm Tr}$\,\{H\Pi \}$ and
this would be a candidate for the average value of $H$ on $|f_{0}>$. In
fact, with the help of some generalized spectral decompositions \cite{21}
for the Hamiltonian in terms of the Gamow vectors and the generalized
eigenvectors of $H$ with eigenvalues in the continuous spectrum of $H$, we
can define this trace in such a way that {\rm Tr}$\,\{H\Pi \}=<\widetilde{f}%
_{0}|H|f_{0}>$ \cite{21}, thus

\begin{equation}
<\widetilde{f}_{0}|H|f_{0}>=z_{R}<\widetilde{f}_{0}|f_{0}>=z_{R}  \label{12}
\end{equation}
Yet this result cannot be acceptable from the physical point of view. Due to
the time-energy uncertainty principle we cannot measure {\it simultaneously}
the real part of $z_{R}$, which is the resonant energy, and its imaginary
part, which is proportional to the inverse of the half life. Thus, $z_{R}$
cannot be the {\it average} of any measurement process and cannot be
accepted as the energy average. Also, from these considerations, we conclude
that the energy average of a Gamow vector, if it can be defined, should be
real.

\subsection{The averages are real in the interpretation of Bohm.}

This point of view is based in the idea that it is possible to construct a
rigged Hilbert space, in which the Gamow vector is a vector in the Hilbert
space, under the following conditions:

1. The continuous spectrum of $H$ is the whole real axis,

2. $H$ is not self adjoint (although it is still symmetric, i.e., $<\phi
|H\psi >=<H\phi |\psi >$ for all $\phi $ and $\psi $ in the domain of $H$),
and,

3. from the point of view of the Hilbert space, the Gamow vector is not in
the domain of $H$, but the action of $H$ on the Gamow vector is well defined
in the dual space (that includes the Hilbert space).

The spaces of analytic functions on a half plane that we are using here are
spaces of Hardy functions \cite{22}\cite{23}\cite{24}\cite{25}. These
functions are determined by their boundary values on the positive semi-axis $%
R^{+}=[0,\infty )$. Moreover, we have an explicit formula to recover their
values on the half plane (including the negative semi-axis, $R^{-}=(-\infty
,0]$) from their values on the positive semi-axis \cite{18}. If we denote by 
${\cal H}_{\pm }^{2}$, the spaces of Hardy functions on the upper (+) and
lower (-) half planes, and by ${\cal H}_{\pm }^{2}|_{R^{+}}$ the spaces of
the restrictions of the functions of ${\cal H}_{\pm }^{2}$ on the positive
semi-axis, there is a one to one mapping $\theta _{\pm }$ \cite{BOOKBG} such
that 
\begin{equation}
\theta _{\pm }\;{\cal H}_{\pm }^{2}\longmapsto \left. {\cal H}_{\pm
}^{2}\right| _{R^{+}}  \label{13}
\end{equation}
As a matter of fact, as our spaces of analytic functions, we take certain
regular subspaces of ${\cal H}_{\pm }^{2}$. Let $S$ be the space of all
functions from $R$ to $C$ which are differentiable to all orders and that
vanish at $\pm \infty $ faster than the inverse of any polynomial (Schwartz
space). Then, let us consider the spaces $\Psi _{\pm }={\cal H}_{\pm
}^{2}\cap S$. We have the following relation between $\Psi _{\pm }$ and $%
\Phi _{\pm }$ 
\begin{equation}
\theta _{\pm }\;\Psi _{\pm }\longmapsto \Delta _{\pm };\;\;\;V_{\pm }=\Omega
_{\pm }U^{-1}\;\Delta _{\pm }\longmapsto \Phi _{\pm }\;,  \label{14}
\end{equation}
or, equivalently, 
\begin{equation}
\Phi _{\pm }=V_{\pm }\theta _{\pm }\Psi _{\pm }  \label{15}
\end{equation}
The spaces of Hardy functions ${\cal H}_{\pm }^{2}$ are Hilbert spaces as
subspaces of $L^{2}(R)$. Therefore, the norms in ${\cal H}_{\pm }^{2}$ and
in $L^{2}(R)$ coincide. The mappings $\theta _{\pm }$ are one to one
transformations from ${\cal H}_{\pm }^{2}$ into $L^{2}(R^{+})$. In this
sense $\theta _{\pm }$ are not unitary, from $L^{2}(R)$ onto $L^{2}(R^{+})$,
because for any $\varphi _{\pm }(E)\in {\cal H}_{\pm }^{2}$ 
\begin{equation}
||\theta _{\pm }\varphi _{\pm }||_{L^{2}(R^{+})}=\int_{0}^{\infty }|\varphi
_{\pm }(E)|^{2}\,dE<\int_{-\infty }^{\infty }|\varphi _{\pm
}(E)|^{2}\,dE=||\varphi _{\pm }||_{L^{2}(R)}<\infty   \label{16}
\end{equation}
In fact, $\varphi _{\pm }$ are boundary values of analytic functions and
cannot be zero on the negative semi-axis unless they vanish identically.

Now, we can construct a new rigged Hilbert space which is given by $\Psi
_{\pm }\subset {\cal H}_{\pm }^{2}\subset \Psi _{\pm }^{\times }$. The
mappings $\theta _{\pm }$, induce two one to one mappings, $\theta _{\pm
}^{\times }$, from $\Psi _{\pm }^{\times }$ onto $\Delta _{\pm }^{\times }$,
by means of the identity 
\begin{equation}
<\theta _{\pm }\varphi _{\pm }|\theta _{\pm }^{\times }G_{\pm }>=<\varphi
_{\pm }|G_{\pm }>,  \label{17}
\end{equation}
where $\varphi _{\pm }\in \Psi _{\pm }$ and $G_{\pm }\in \Psi _{\pm
}^{\times }$. The mappings $\theta _{\pm }^{\times }$ are not extensions of $%
\theta _{\pm }$ because of the non-unitary of $\theta _{\pm }$. On the other
hand, the unitary of $V_{\pm }$ allows us to extend them into the dual
spaces by means of a similar formula:, i.e. if $\phi _{\pm }\in \Delta _{\pm
}$ then $V_{\pm }\phi _{\pm }\in \Phi _{\pm }$ and if $F_{\pm }\in \Delta
_{\pm }^{\times }$ then $V_{\pm }F_{\pm }\in \Phi _{\pm }^{\times }$ such
that 
\begin{equation}
<V_{\pm }\phi _{\pm }|V_{\pm }F_{\pm }>=<\phi _{\pm }|F_{\pm }>  \label{18}
\end{equation}

Thus, to any $\varphi \in \Phi _{\pm }$ corresponds an analytic function $%
\phi _{\pm }(E)\in \Psi _{\pm }$ and $\phi _{\pm }(E)=\theta _{\pm
}^{-1}V_{\pm }^{-1}\varphi _{\pm }$. Therefore, the Gamow vectors can be
represented as vectors in $\Psi _{\pm }$ as $(\theta _{+}^{\times
})^{-1}V_{+}^{-1}|f_{0}>$ and $(\theta _{-}^{\times })^{-1}V_{-}^{-1}|%
\widetilde{f}_{0}>$. However these formulas are unpractical to obtain the
Gamow vectors in the new representation. In order to find them we shall use
the definition of $|f_{0}>$ and $|\widetilde{f}_{0}>$ and the Titchmarsh
theorem \cite{22}\cite{23}\cite{24}\cite{25} on Hardy functions. Then, we
have 
\begin{equation}
<\varphi _{+}|f_{0}>=\phi _{+}^{\#}(z_{R}^{*})=-\frac{1}{2\pi i}%
\int_{-\infty }^{\infty }\frac{\phi _{+}^{\#}(E)}{E-z_{R}^{*}}\,dE
\label{19}
\end{equation}
and 
\begin{equation}
<\varphi _{-}|\widetilde{f}_{0}>=\phi _{-}^{\#}(z_{R})=\frac{1}{2\pi i}%
\int_{-\infty }^{\infty }\frac{\phi _{-}^{\#}(E)}{E-z_{R}}\,dE  \label{20}
\end{equation}
These formulas imply that 
\begin{equation}
(\theta _{+}^{\times })^{-1}V_{+}^{-1}|f_{0}>=\frac{-1}{2\pi i}\,\frac{1}{%
E-z_{R}^{*}}  \label{21}
\end{equation}
and 
\begin{equation}
(\theta _{-}^{\times })^{-1}V_{-}^{-1}|\widetilde{f}_{0}>=\frac{1}{2\pi i}\,%
\frac{1}{E-z_{R}}  \label{22}
\end{equation}

In this representation the Gamow vectors are square integrable, i.e., they
belong to the Hilbert space $L^{2}(R)$. Therefore, we can define brackets
and scalar products between them. We can also define energy averages on
these vectors. This is, however, not an easy task. First of all, we must
observe that in the new representation, the Hamiltonian $H$ is given by ${%
\widehat{E}}={V}_{\pm }^{-1}H{V}_{\pm }$, where ${V}_{\pm }=V_{\pm }\theta
_{\pm }^{-1}$. It is easy to show that ${\widehat{E}}\phi _{\pm }(E)=E\phi
_{\pm }(E)$, i.e., the multiplication operator on $L^{2}(R)$. This is why we
do not add subscripts in ${\widehat{E}}$. On ${\cal H}_{\pm }^{2}$, ${%
\widehat{E}}$ is still symmetric but it is not self adjoint (has different
deficiency indices on ${\cal H}_{\pm }^{2}$). Its spectrum is purely
continuous, simple and coincides with $R$. The definition of energy averages
in this representation is hampered by the fact that 
\begin{equation}
{\widehat{E}}\frac{1}{E-z_{R}}=\frac{E}{E-z_{R}}  \label{23}
\end{equation}
is not square integrable. This only means that the function $(E-z_{R})^{-1}$
does not belong to the domain of the multiplication operator ${\widehat{E}}$%
. However, since ${\widehat{E}}$ can be extended by continuity to $\Psi
^{\times }$, the identity (\ref{23}) makes sense.

The representations (Eqs (\ref{21})) and (Eqs (\ref{22})) of the Gamow
vectors, when properly normalized, could be used now to define a mean value
of the energy for these states. The normalization that we are going to use
is the Hilbert space normalization, i.e., if we call 
\begin{equation}
\psi ^{D}=\alpha \,\frac{1}{E-z_{R}}\;;\;\;\;\psi ^{G}=\alpha \,\frac{1}{%
E-z_{R}^{*}}\;,  \label{24}
\end{equation}
then 
\begin{equation}
{||\psi ^{D}||}^{2}=\alpha ^{2}\,\int_{-\infty }^{\infty }\frac{dE}{{%
(E-z_{R})}^{2}+{(\Gamma /2)}^{2}}={\alpha }^{2}\pi  \label{25}
\end{equation}
Therefore, $||\psi ^{D}||=||\psi ^{G}||=1$ if $\alpha =1/\sqrt{\pi }$.

Now, let us define the mean value of the energy on the decaying Gamow vector
as 
\begin{equation}
<\psi ^{D}|{\widehat{E}}|\psi ^{D}>  \label{26}
\end{equation}
and let us evaluate its value. Since ${\widehat{E}}\,\psi ^{D}(E)=E\,\psi
^{D}(E)$, we have that 
\begin{equation}
<\psi ^{D}|{\widehat{E}}|\psi ^{D}>=\frac{1}{\pi }\int_{-\infty }^{\infty }%
\frac{1}{E-z_{R}^{*}}\;\frac{E}{E-z_{R}}\,dE=\frac{2}{\pi \Gamma }%
\int_{-\infty }^{\infty }\frac{E\,dE}{\left( \frac{E-E_{R}}{\Gamma /2}%
\right) ^{2}+1}  \label{27}
\end{equation}
The change of variables 
\begin{equation}
x=\frac{E-E_{R}}{\Gamma /2}  \label{28}
\end{equation}
transforms the last integral in (\ref{27}) into 
\begin{equation}
\frac{E_{R}}{\pi }\,\int_{-\infty }^{\infty }\frac{dx}{x^{2}+1}\,+\,\frac{%
\Gamma }{2\pi }\int_{-\infty }^{\infty }\frac{x\,dx}{x^{2}+1}  \label{29}
\end{equation}
The first integral in (\ref{29}) has the value $\pi $. The second admits a
Cauchy principal value equal to zero. Thus, we find 
\begin{equation}
<\psi ^{D}|{\widehat{E}}|\psi ^{D}>=E_{R}  \label{30}
\end{equation}
and 
\begin{equation}
<\psi ^{G}|{\widehat{E}}|\psi ^{G}>=E_{R}  \label{31}
\end{equation}

We see that this definition of the energy average of Gamow vectors gives the
same real value for both Gamow vectors and coincides with the resonant
energy. In addition, due to the adopted normalization, we have that $<\psi
^{D}|\psi ^{D}>=<\psi ^{G}|\psi ^{G}>=1$. Furthermore 
\begin{equation}
<\psi ^{G}|\psi ^{D}>=\frac{1}{\pi }\,\int_{-\infty }^{\infty }\frac{dE}{%
(E-z_{R})^{2}}=0  \label{32}
\end{equation}
Analogously, $<\psi ^{D}|\psi ^{G}>=0$.

\subsection{The averages are real in Berggren's interpretation.}

Berggren's approach to the mean value of the Hamiltonian on a Gamow state 
\cite{12} can be formulated very similarly to Bohm's. Following \cite{26}%
\cite{Bollini96}, we shall not use Hardy functions to construct our Gelfand
triplets. Instead, we consider here another triplet ${\widetilde{\xi }%
\subset {\cal H}\subset {\widetilde{\xi }}^{\times }}$ for which the space ${%
\widetilde{\xi }}^{\times }$ consists of tempered ultra-distributions. A
simple definition of these objects can be found in \cite{26}\cite{Bollini96}
and a complete account in \cite{27}\cite{Hasumi61}. Vectors in $\widetilde{%
\xi }$ are entire analytic functions. Vectors in ${\widetilde{\xi }}^{\times
}$ are represented by pairs of analytic functions on the open upper and
lower half planes, respectively. If we call $\psi _{u}(z)$ and $\psi _{l}(z)$
these functions, we can write 
\begin{equation}
\psi (E)=\psi _{u}(E+i0)-\psi _{l}(E-i0),  \label{33}
\end{equation}
where $\psi _{u}(E+i0)$ and $\psi _{l}(E-i0)$ represent the boundary limits
of $\psi _{u}(z)$ and $\psi _{l}(z)$ on the real axis, respectively. For any 
$\psi \in {\widetilde{\xi }}^{\times }$ and any $z\in C-R$, we have 
\begin{equation}
\psi (z)=\pm \frac{1}{2\pi i}\int_{-\infty }^{\infty }\frac{\psi (E)}{E-z}%
\,dE  \label{34}
\end{equation}
were we use in (\ref{34}) the sign $+$ or $-$ for $z$ on the upper or lower
half plane, respectively, 
\begin{equation}
\psi (z_{R})=\frac{1}{2\pi i}\int_{-\infty }^{\infty }\frac{\psi (E)}{E-z_{R}%
}\,dE\;,\;\;\;\psi (z_{R}^{*})=-\frac{1}{2\pi i}\int_{-\infty }^{\infty }%
\frac{\psi (E)}{E-z_{R}^{*}}\,dE  \label{35}
\end{equation}
Now, we have a similar scheme to that presented in the previous section. The
total Hamiltonian $H$ is represented by the multiplication operator on $%
\widetilde{\xi }$ and it can be extended as a continuous operator into the
dual $\widetilde{\xi }^{\times }$. The nuclear spectral theorem guarantees
the existence of a complete set of eigenvectors $|E>$ of $H$ \cite{28}. As
eigenvectors, they obey the identity $H|E>=E|E>$. The Gamow vectors are now
defined as\footnote{%
The decaying Gamow vector is denoted in [35-36] as $|E_{G}^{*}>$ and the
growing Gamow vector as $|E_{G}>$.} 
\begin{equation}
|f_{0}>=-\frac{1}{2\pi i}\int_{-\infty }^{\infty }\frac{|E>\,dE}{E-z_{R}}%
\;,\;\;\;|\widetilde{f}_{0}>=\frac{1}{2\pi i}\int_{-\infty }^{\infty }\frac{%
|E>\,dE}{E-z_{R}^{*}}.  \label{36}
\end{equation}
Within this scheme, the mean value of the Gamow vectors is defined as in the
previous section and it gives the same results\footnote{%
In both cases, we can define the probability distribution associated to the
Gamow states and it is given by (for the decaying Gamow vector) $%
P(E)=|<E|f_{0}>|^{2}=\frac{1}{\pi }\frac{\Gamma }{(E-E_{R})^{2}+\Gamma ^{2}}$%
. The same expression is found for the growing Gamow vector.}.

The coincidence between this last result for the mean value of the Gamow
states and Bohm's one, presented in the last sub-section, comes from a
re-interpretation of Berggren's definition given in \cite{12}. In fact, for
a spherically symmetric potential and for an arbitrary value of the angular
momentum $l$, we can write the normalized decaying Gamow vector as 
\begin{equation}
|f_{0}>=i\,\sqrt{\frac{2\Gamma }{\pi }}\int_{0}^{\infty }\sqrt{\frac{k}{m}}\;%
\frac{|k,{\hat{k}},l>}{E({\bf k})-z_{R}}\,dk,  \label{37}
\end{equation}
where, $k=|{\bf k}|$, $E({\bf k})=k^{2}/2m$ and ${\hat{k}}$ is the unit
vector in the direction of ${\bf k}$. For the growing Gamow vector, we have 
\begin{equation}
|\widetilde{f}_{0}>=-i\,\sqrt{\frac{2\Gamma }{\pi }}\int_{0}^{\infty }\sqrt{%
\frac{k}{m}}\;\frac{|k,{\hat{k}},l>}{E({\bf k})-z_{R}^{*}}\,dk.  \label{38}
\end{equation}

Now, let $A$ be an arbitrary observable. We can define the mean value of $A$
on $|f_{0}>$ as 
\begin{equation}
<f_{0}|A|f_{0}>=\frac{2\Gamma }{\pi }\sum_{l,l^{\prime }}\int_{0}^{\infty
}dk\int_{0}^{\infty }dk^{\prime }\,\frac{\sqrt{kk^{\prime }}}{m}\;\frac{%
<k^{\prime },{\hat{k}}^{\prime },l^{\prime }|A|k,{\hat{k}},l>}{(E({\bf %
k^{\prime }})-z_{R})(E({\bf k})-z_{R}^{*})}\;.  \label{39}
\end{equation}

If we replace $A$ by $H$ we obtain, straightforwardly, the value $E_{R}$ for
this average. Instead, Berggren defines \cite{12} this mean value as ${\rm %
Real}<\widetilde{f}_{0}|A|f_{0}>$. As a matter of fact, one can easily show
that 
\begin{equation}
<f_{0}|A|f_{0}>={\rm Real}\{<\widetilde{f}_{0}|A|f_{0}>\}+o(\Gamma ^{2}),
\label{40}
\end{equation}
which means that Berggren's approximation coincides with Bohm's to the first
order in $\Gamma $.

\section{Conclusions}

In this paper we have compared different definitions of Gamow vectors. We
have shown the equivalence between Bohm's and Berggren's definitions of the
mean value of the Hamiltonian on a resonant state. Our main result,
concerning this equivalence, is the realization of the average value of the
Hamiltonian on a resonance as a real function which depends on both the real
and the imaginary parts of the complex energy. This result is supported,
mathematically, by a proper treatment of Gamow vectors in a rigged Hilbert
space.

\section{Acknowledgements}

This work has been partialy supported by the Junta de Castilla y Le\'{o}n,
project number CO2/199, the Spanish DGICYT PB 95-0719, the Intercampus
Programme, the CONICET (Argentina) and the University of La Plata
(Argentina).


\begin{references}
\bibitem{liotta96}  R. J. Liotta, E. Maglione, N. Sandulescu, T. Vertse,
Phys. Lett. B {\bf 367} (1996) 1.

\bibitem{REP}  R. G. Lovas, R. J. Liotta, A. Insolia, K. Varga and D. S.
Delion, Phys. Rep. {\bf {294}} (1998) 265.

\bibitem{GAMOW}  G. Gamow, Z. Phys. {\bf {51}} (1928) 204.

\bibitem{TORE}  T. Berggren, Nucl. Phys. {\bf {A 109}} (1968) 265.

\bibitem{LIOTTA}  P. Curuchet, T. Vertse and R. J. Liotta, Phys. Rev. {\bf {%
C 39}} (1989) 1020.

\bibitem{VERTSE}  B. Gyarmati and T. Vertse; Nucl. Phys. {\bf {A 160}}
(1971) 523.

\bibitem{BOOKBG}  A. Bohm and M.~Gadella, {\it Dirac Kets, Gamow Vectors and
Gel'fand Triplets}, Lecture Notes in Physics, vol. {\bf 348} (Springer,
Berlin, 1989).

\bibitem{1}  J. von Neumann, {\it Mathematical Foundations of Quantum
Mechanics}, (Princeton University Press 1955).

\bibitem{2}  M. L. Goldberg and K. M. Watson, {\it Collision Theory}, Wiley,
New York (1964).

\bibitem{3}  R. G. Newton, {\it Scattering Theory of Waves and Particles},
(Springer Verlag, Berlin, 1982).

\bibitem{4}  A. Bohm, {\it Quantum Mechanics Foundations and Applications}
1st Ed. (Springer, New York, 1979); 3rd Ed. (1993).

\bibitem{5}  P. Exner, {\it Open Quantum Systems and Feynman Integrals},
Reidel, Dordrech (1985).

\bibitem{7}  A. Bohm, M. Gadella, S. Wicramasekara, {\it Some Little Things
About Rigged Hilbert Spaces and Quantum Mechanics and All That.}, in {\it %
Generalized Functions, Operator Theory and Dynamical Systems}, CRC Research
Notes in Mathematics, Chapman and Hall, London (1999).

\bibitem{8}  N. Nakanishi, Progr. Theor. Phys. {\bf 19} (1958) 607.

\bibitem{9}  M. Castagnino, R. Laura, Phys. Rev. A, {\bf 56} (1996) 108.

\bibitem{10}  M. Castagnino, E. Gunzig, Int. J. Theor. Phys., {\bf 38}
(1999) 47.

\bibitem{12}  T. Beggren, Phys. Lett. B, {\bf 373} (1996) 1.

\bibitem{13}  L. Fonda, G.C. Ghirardi, A. Rimini, Rep. Prog. Phys., {\bf 41}
(1978) 587.

\bibitem{14}  A. Bohm, J. of Math. Phys. {\bf 21} (1980) 1040.

\bibitem{BohmA81}  A. Bohm, J. of Math. Phys. {\bf 22} (1981) 2813.

\bibitem{15}  M. Gadella, J. of Math. Phys. {\bf 24} (1983) 1462.

\bibitem{Gadella83}  M. Gadella, J. of Math. Phys. {\bf 24} (1983) 2142.

\bibitem{Gadella84}  M. Gadella, J. of Math. Phys. {\bf 25} (1984) 2481.

\bibitem{16}  A. Bohm, I. Antoniou, P. Kielanowski, J. Math. Phys., {\bf 36}%
, (1995) 2593.

\bibitem{17}  H.M. Nussenzveig, {\it Causality and Dispersion Relations},
Academic, New York (1972).

\bibitem{18}  C. van Winter, J. Math. Anal., {\bf 47}, (1974) 633.

\bibitem{19}  I. E. Antoniou, M. Gadella, G. P. Pronko, J. Math. Phys., {\bf %
39} (1998) 2459.

\bibitem{A}  T. Petroski, I. Prigogine, S. Tasaki, Physica A, {\bf 173}
(1991) 175.

\bibitem{20}  I. E. Antoniou and I. Prigogine, Physica A, {\bf 192}, (1993)
443.

\bibitem{21}  M. Gadella, Int. J. Theor. Phys., {\bf 36}, (1997) 2271.

\bibitem{22}  K. Hoffmann, {\it Banach Spaces of Analytic Functions},
(Prentice Hall, New Jersey, 1962).

\bibitem{23}  P. L. Duren, {\it Theory of $H^{p}$ Spaces}, (Academic Press,
New York, 1970).

\bibitem{24}  P. Koosis, {\it The Logarithmic Integral}, (Cambridge, UK,
1990).

\bibitem{25}  E. C. Titchmarsh, {\it Introduction to the Theory of Fourier
Integrals}, (Clarendon Press, Oxford, UK, 1937).

\bibitem{26}  C. G. Bollini, O. Civitarese, A. L. De Paoli, M. C. Rocca,
Phys. Lett. {\bf B 382} (1996) 205.

\bibitem{Bollini96}  C. G. Bollini, O. Civitarese, A. L. De Paoli, M. C.
Rocca, J. of Math. Phys. {\bf 37} (1996) 4235.

\bibitem{27}  J. Sebasti\~{a}o e Silva, Math. Ann., {\bf 136} (1958) 38.

\bibitem{Hasumi61}  M. Hasumi, T${\rm {\hat{o}}}$hoku Math. J. {\bf 13}
(1961) 94.

\bibitem{28}  I. M. Gel'fand and N. Ya. Vilenkin, Generalized Functions {\bf %
Vol. 4}. Academic Press (1964).
\end{references}
\end{document}